 \def\Title{Reply to ``The three-box paradox revisited'' by Ravon and Vaidman}
 \def\arXiv{quant-ph/0703004}
 \def\Abstract{I reply to Ravon and Vaidman's criticism (\emph{J. Phys. A} {\bf 40}
  2873 (2007); quant-ph/0606067) of my classical implementation of a Three-Box system as a card-game.}

\documentclass[aps,rmp,nobibnotes,nofootinbib,onecolumn,twoside,noshowpacs,fleqn,draft]{revtex4}
 \usepackage[tbtags,sumlimits]{amsmath}
 \usepackage{amssymb}

 \def\thesection{\arabic{section}}
 \def\thesubsection{\arabic{section}.\arabic{subsection}}
 \def\thesubsubsection{\arabic{section}.\arabic{subsection}.\arabic{subsubsection}}
 \def\theparagraph{\arabic{section}.\arabic{subsection}.\arabic{subsubsection}.%
                     \arabic{paragraph}}
 \def\thesubparagraph{\arabic{section}.\arabic{subsection}.\arabic{subsubsection}.%
                     \arabic{paragraph}.\arabic{subparagraph}}
 \makeatletter
 \def\p@section{}
 \def\p@subsection{}
 \def\p@subsubsection{}
 \def\p@paragraph{}
 \def\p@subparagraph{}

 \def\@hangfrom@section#1#2#3{\@hangfrom{#1}{\textrm{\large#2}}{\textrm{\large#3}}}%
 \def\@hangfrom@subsection#1#2#3{\@hangfrom{#1}{\textrm{#2}}{\textrm{#3}}}%
 \def\@hangfrom@subsubsection#1#2#3{\@hangfrom{#1}{\textrm{#2}}{\textrm{#3}}}%
 \def\frontmatter@setup{\normalfont\raggedright}
 \makeatother

 \usepackage{xspace}

 \newcommand{\ie}{i.e.,\  }
 \newcommand{\eg}{e.g.,\  }
 \newcommand{\cf}{cf.\ }
 \newcommand{\Cf}{Cf.\ }

 \newcommand{\ket}[1]{\ensuremath{\vert\,{#1}\,\rangle}}

 \newcommand{\Face}{\ensuremath{\text{\textsl{Face}}}\xspace}
 \newcommand{\Suit}{\ensuremath{\text{\textsl{Suit}}}\xspace}
 \newcommand{\K}{\ensuremath{\text{\textsf{K}}}\xspace}
 \newcommand{\Q}{\ensuremath{\text{\textsf{Q}}}\xspace}
 \newcommand{\J}{\ensuremath{\text{\textsf{J}}}\xspace}
 
 \renewcommand{\S}{\ensuremath{\text{\textsf{S}}}\xspace}
 \renewcommand{\H}{\ensuremath{\text{\textsf{H}}}\xspace}
 \newcommand{\D}{\ensuremath{\text{\textsf{D}}}\xspace}

\begin{document}
 \makeatletter
 \def\ps@titlepage{%
   \renewcommand{\@oddfoot}{}%
   \renewcommand{\@evenfoot}{}%
   \renewcommand{\@oddhead}{\emph{J.~Phys.~A} {\bf 40}(11) 2883-2890 (2007)\hfill\arXiv}
   \renewcommand{\@evenhead}{}}
 \makeatother

\title[Kirkpatrick -- \Title] 
      {\Title} 

\author{K.~A.~Kirkpatrick}
\affiliation{New Mexico Highlands University, Las Vegas, New Mexico 87701\\ \textnormal{\texttt{kirkpatrick@physics.nmhu.edu}}}
\begin{abstract}
 \Abstract
\end{abstract}


 \maketitle
 \makeatletter\markboth{\hfill\@shorttitle\hfill}{\hfill\@shorttitle\hfill}\makeatother
 \pagestyle{myheadings}

\section{Introduction}
The Three-Box ``paradox''%
\footnote{%
Presented in careful detail in Appendix \ref{A:ThreeBox}
} %
was introduced by \citet*{AharonovV91} (henceforth ``AV''); it has been engaged in controversy%
\footnote{%
For example, these two exchanges: \citet{Vaidman99a}, \citet{Kastner99}, \citet{Vaidman99b}, and  \citet{AharonovV03}, \citet{Kastner04}.
} %
regarding its significance and meaning ever since. Vaidman (\eg (\citeyear{Vaidman99a}), (\citeyear{Vaidman99b})) has presented it as strange even within the ordinary strangeness of quantum mechanics. 

I have argued \citep{Kirkpatrick03b} for what I see as the quite ordinary, non-paradoxical nature of the system. I presented, as an illustration of its ordinariness, a classical Three-Box-like system implemented as a playing-card game. In the current issue of this journal \citet{RavonVaidman07} argue to reject this classical example; this article is my reply.

\begin{equation}
\text{\parbox{0.85\textwidth}{\textbf{The paradox} in the Three-Box experiment is that at a particular time we can claim that a particle is in some sense both with certainty in one box, $A$, and with certainty in another box, $B$. Now, if a particle is certainly in $A$, then it is certainly not in $B$, and vice versa. Therefore, if a single particle is both certainly in $A$ and certainly in $B$ we have a paradox \dots. In the Three-Box experiment, the particle is \emph{certain} to be found in $A$ if searched for in $A$, and \emph{certain} to be found in $B$ if searched for in $B$ instead.}}
\end{equation}
\noindent This description, from Ravon and Vaidman, is typical: one particle is to be found in each of several boxes with certainty. It is reasonable to assert, though, that the ``paradox'' is not specific to boxes and a particle, rather that it lies in the apparent simultaneous certainty of two disjoint events.

\section{The Card-Game analogy}
As Ravon and Vaidman acknowledge, ``The existence of a classical system which can perform the Three-Box task would remove the paradox.'' In my (\citeyear{Kirkpatrick03b}) I described a system constructed from ordinary playing cards which exactly performed this task. 

The classical Three-Box ``paradox'' system is expressed in terms of a partial deck of playing cards and an internal mechanism (set of rules) for observation of the value of the two variables \Suit and \Face. Ravon and Vaidman have re-presented this game quite clearly in their Sec.~4 and Fig.~1, so here I present merely a brief sketch of the system and its ``physics'':

The three boxes correspond to the three values \S, \D, and \H of the variable \Suit.  The values of the variable \Face are \K, \Q, \J.  The deck is separated into two parts, \emph{These} and \emph{Others}.%
\footnote{%
These terms are carryovers from an earlier example in which they had meaning. In the published version of \citep{Kirkpatrick03b} they are called \emph{Active} and \emph{Passive}. Unfortunately an editing error put the current terms into the arXiv version of that paper.
} %
To observe the value of a variable, pick a card at random --- if the variable (\Face or \Suit) being observed is the same as that of the last observation, pick the card from \emph{These}, otherwise pick it from \emph{Others} --- read the card's value of the variable being observed, and place all such cards into \emph{These} and the remainder into \emph{Others}.

One need merely write ``card system'' for ``particle''  and ``$\Suit=\S$ (respectively \D)'' for ``ball in box $A$ (respectively $B$)'' to express the Three-Box system as the card game.

This mechanism yields repeatable observations. Repeatability, in the Three-Box example, is required in order to be certain that the ball, once observed in a box, remains there.  (The properties of systems such as this one are discussed in detail in \citep{Kirkpatrick03a}.) 

In their Abstract, Ravon and Vaidman claim that ``It is shown that Kirkpatrick's card game is analogous to a different game with a particle in three boxes which does not contain paradoxical features.'' In their Introduction, they state ``We analyze Kirkpatrick's card game and show that it does not reproduce the paradoxical features of the original Three-Box experiment,'' and then describe those paradoxical features in the immediately following paragraph (quoted above, in (1)). 

But simply working out the card game example (nicely presented by Ravon and Vaidman in their Fig.~1) makes it clear that my card game is entirely successful at reproducing those so-defined ``paradoxical features of the original Three-Box experiment.'' In the sense stipulated by Ravon and Vaidman,  the paradox has thus been removed.

\section{Ravon and Vaidman's criticism of the card game}

Then what of the claim that my card game was analyzed and found wanting? Well --- it was never analyzed. Instead, my card game was simply replaced by another game, one which lacks my system's ``complications of \dots\ system states and variables.'' Then Ravon and Vaidman claim that in this simplified game ``parts of the system are moved around in such a way that the post-selection becomes impossible whenever the measurement is unsuccessful.''

They then introduce what I will call the Ad Hoc Three-Box system, in which the ball is indeed ``moved around'': If the observer does not find the ball in the first box, it is placed in the third (the post-selection is ``not in third box''). This is a truly \emph{ad hoc} mechanism, designed to guarantee the result; according to Ravon and Vaidman, it is thus ``entirely non-paradoxical.'' 

Absurdly, immediately following their introduction of the Ad Hoc Three-Box system, they announce --- with neither demonstration nor discussion --- that it is to this Ad Hoc Three-Box system, rather than to the Three-Box system itself, that my card game is ``analogous.'' 

Thus was my card game ``analyzed.''

The use of the passive voice in ``parts are moved around'' rhetorically suggests a forcing of the desired result by an agent (as is in fact the case for the Ad Hoc system). But, although examination of how the parts move shows that they do indeed move around ``in such a way that the post-selection becomes impossible whenever the measurement is unsuccessful,'' no agent exists whose goal is satisfied by that motion. The parts move according to the rules of the system (its ``physics''),%
\footnote{%
The distinction is fundamental to physics. For example, it distinguishes the ad hoc solar system models of Ptolemy and Copernicus from the dynamical model of Newton --- Ptolemy moves the planets \emph{ad hoc}, Newton lets them move according to the rules.
} %
rules which are entirely unrelated to (and were constructed in ignorance of) the Three-Box experiment.
  
Further, it is devastating to Ravon and Vaidman's position that their remark, expressed in terms of states rather than parts, applies equally to the original Three-Box quantum example: \emph{The state of the system is modified by the observation in such a way that the post-selection becomes impossible whenever the measurement is unsuccessful}. (\Cf Appendix~\ref{A:ThreeBox}, particularly at footnote \ref{F:modify}.)

\section{The ``Quantum Paradox''}

Ravon and Vaidman introduce the claim that the Three-Box system is a different sort of paradox,%
\footnote{%
Ravon and Vaidman here define ``a true physical paradox'' as ``a prediction of a current physical theory that contradicts experimental results.'' This highly idiosyncratic expression for what ordinarily would be called a ``failure of physical theory'' raises the question of just what they mean by the word ``paradox.''  
} %
a ``quantum paradox'': ``What we mean by ``quantum paradox'' is a phenomenon that \emph{classical} physics cannot explain.'' But this is exactly the definition of a quantum  phenomenon --- ``quantum paradox,'' so defined, is trivially equivalent to \emph{quantum mechanics}! This definition of quantum paradox is empty.

Another comment, although not stated formally as a definition of quantum paradox, seems to define it implicitly:
\begin{quote}
It is precisely because the classical observation in the Three-Box experiment is \emph{non-disturbing} that the experiment cannot be explained by classical physics or accomplished using classical means. This is what causes the Three-Box experiment to be a ``quantum paradox.''
\end{quote}
Unfortunately, Ravon and Vaidman do not define or discuss what they mean by a ``classical'' observation; they do explain that ``spin measurement is genuinely quantum,'' hence \emph{not} classical --- but do not explain what ``genuinely quantum'' means. 

It appears that Ravon and Vaidman have not succeeded in establishing the Three-Box system as a ``quantum paradox,'' nor have they made sense of that term.

\section{Measurement and disturbance}

Ravon and Vaidman state that ``observation in the Three-Box experiment is \emph{non-disturbing},'' while they speak of ``measurement disturbance'' in my card game ``leaving a mark'' which ``encodes'' the outcome of the observation for the post-selection process to ``read.'' They say 
\begin{quote}
[in Kirkpatrick's card game] the ``observation'' includes leaving a mark that is later ``read' in the post-selection process. The observation in the original Three-Box experiment, which consists of opening a box, does not leave such a mark in the framework of classical physics.
\end{quote}

But this is incorrect.  Ravon and Vaidman mean to distinguish my (classical) card game from the (quantum mechanical) Three-Box system, but in no sense does either leave a ``mark'' to be ``read.'' In both, observation leaves the system in such a state that post-selection is impossible whenever a detection occurred.

To make this clear, first let us consider a system which actually is disturbed by observation --- the Leifer-Spekkens box (\citet{LeiferSpekkens05}, presented by Ravon and Vaidman in their Sec.~5). The two variables of this system are the location of the ball in the front or rear half, and the location of the ball in the left or right half.  To observe if the ball is in, say, the left half, the box is split in half, and the left half is shaken. A rattling sound confirms its presence, and, of course, randomizes the position of the ball front to rear --- a disturbance due to observation.

However, in the Leifer-Spekkens box measurement needn't disturb the variables' values.  Rather than shaking the box, it could be swiftly turned to a vertical position, measuring the time for the sound of the ball hitting the bottom end, directly telling not only that it is present in the left half, but whether it was in the front or rear.  By tilting the box, the ball could be returned to its original position, so the observation does not disturb the system. (Alternatively, we might avoid all this trouble and just X-ray the box to locate the ball!)

Measurement disturbance and non-disturbance are possible only because the ball actually \emph{has}, prior to observation, a position to be disturbed --- if it did not have a value prior to observation, there would be nothing to disturb. That, I believe, is the case in quantum mechanics; it is certainly the case in my card game: In my card game there is no possibility of an X-ray-like examination to determine a pre-observation value, because there \emph{is} no pre-observation value. If the system has been prepared say as a value of \Face, or if the most recent observation has been of \Face, then \Suit has no value. The most thorough examination of the contents of \emph{These} and \emph{Others} will neither suggest a current hidden value of \Suit nor tell us the value which will occur in an upcoming observation of \Suit.  Thus there is nothing to disturb, and hence no meaning in speaking of ``disturbing'' or ``non-disturbing'' observations.

Of course, there is an \emph{effect} of measurement on the state of the system (``pick a card from \emph{Others} and place all cards with its value of \Suit into \emph{These} and all others into \emph{Others}''), but this is exactly equivalent to the effect of measurement on a quantum system such as the Three-Box: There, the state of the system changes from the pure superposition of the three box states to a mixture of the observed box's state and a superposition of the other two box states.  Observation in the card game example ``disturbs'' exactly as in a quantum system.

Thus, as I claimed above, in neither my card game nor the Three-Box system does observation leave a mark to be read by an \emph{ad hoc} post-selector; rather, in both  the observation changes the statistics of the system in such a way as to make post-selection impossible if a detection occurred. As stated in footnote \ref{F:modify}, \emph{The state of the system is modified by the observation in such a way that whenever the box is empty, the post-selection becomes impossible}.

\section{Discussion and miscellaneous comments}

\subsection{Straw-man argument}
In their criticism of my card game example, Ravon and Vaidman make a ``straw man'' error of rhetoric: In paraphrase, they say ``We are going to show Kirkpatrick's model is incorrect. To do that, we will simplify his model and then show that this simplified model is incorrect. QED.'' Such an approach can be correct only after carefully showing that failure of the simplified model implies failure of the original --- but Ravon and Vaidman made no such effort.

The complex rules of the mechanism of my card game system are there for good reason: Measurements in my system are \emph{repeatable} (in von Neumann's sense), and this system's several variables are \emph{incompatible}.%
\footnote{%
I mean incompatibility in the quantitative sense of \citet{Luders51} or \citet{Davies76}, rather than merely the informal sense of not being simultaneously measurable.
} %
Repeatability allows us to ``look in the box again'' to assure ourselves that the ball is really still there --- necessary to the very meaning of ``the ball is in the box.''  Incompatibility is a critical aspect of the quantum Three-Box --- if the initial or final states were box states (\ie  compatible with looking in the box), not only would there be no ``paradox,'' the results would be entirely ``classical.'' They must be states of a variable not compatible with the box states. 

To ``consider Kirkpatrick's game without the complications of defining \dots\ system states and variables,'' as Ravon and Vaidman do in their ``Simplification'' (their Fig.~2), is to destroy it. There is no possibility of incompatibility in the Simplified model. Furthermore, the Simplified model takes \emph{non}-repeatability to the extreme: In Three-Box terms, after the ball has been found in box~1, if we look again in box~1 it definitely \emph{won't} be there.

\subsection{The Three-Box system is an ordinary quantum system}
Although this exchange centers on my classical ``three-box''-like card game, I believe the clearest realization of the non-paradoxical nature of the Three-Box system (and the best understanding of the system itself) comes from noting that the Three-Box system is identical in quantum formulation to a particular three-slit Young apparatus \citep{Kirkpatrick03b}. (For the reader's convenience I describe that setup in Appendix~\ref{A:ThreeSlit}.)

The physical significance of the pre- and post-selection states of the Three-Box system is entirely obscure when expressed in terms of particles in boxes; we have no idea what those states mean physically, nor how to accomplish them. Furthermore, we have no insight into the destructive interference at the final detector, so we have no insight into the meaning of the ``paradox.'' By restating the Three-Box system as the three-slit Young apparatus, the ``paradox'' is seen to reduce to the interesting but ordinary phenomenon of destructive interference in a \emph{two}-slit Young apparatus, nothing more.

I note that Ravon and Vaidman do not discuss, or even mention, this identity of the Three-Box system with the three-slit Young apparatus, nor did Vaidman in our earlier correspondence.

\subsection{Miscellany}
Ravon and Vaidman suggest that I ``have the following idea'': ``If disturbance is understood to be inherent to measurement, then the difficulty with regard to the Three-Box experiment is removed.''  But this completely misunderstands my position, which is that there is no ``difficulty'' to be ``removed''; the quantitatively identical three-slit form shows that the Three-Box system is perfectly ordinary quantum mechanics, while the behaviorally equivalent card-game example shows that the ``paradoxical'' behavior doesn't even require quantum mechanics.

Ravon and Vaidman comment ``Contrary to the supposedly analogous games, a successful post-selection in the Three-Box experiment is possible even with no intermediate measurement (\ie if none of the boxes are opened).'' That is, in the Three-Box example the pre- and post-selection states are not disjoint. I cannot see the point of this criticism --- nothing in the Three-Box discussion calls on this property --- but in any case it does not apply to my card game: The  ``paradox'' also appears if we replace the example's initial state of $\Q$ with the state $\neg\J$, which is not disjoint from the post-selection \K.%
\footnote{%
The state $\neg\J$ is represented as QS, QD and (2)KH in  \emph{These}, and  JS and JD in \emph{Others} --- obviously not disjoint from $\K$. It is easy to see that the middle row of Ravon and Vaidman's Fig.~1 is the same for the initial state $\J$ as it was for the original example's $\Q$, so the ``paradox'' remains.
} %

Ravon and Vaidman say of observation in the Three-Box system that it is ``just observation'' with no ``additional actions,'' that it ``consists of opening a box''; they call it ``classical,'' though the Three-Box system is explicitly quantum-mechanical. It appears that Ravon and Vaidman treat the boxes as actual classical boxes which can be looked into without effect, permitting the claim of ``paradox.'' But treating the observation as classical is obviously inconsistent with the analysis of the system (as described in Appendix~\ref{A:ThreeBox}, and in AV as well as in Ravon and Vaidman), where the act of observation breaks the coherence between the states ``particle seen in opened box'' and ``particle not seen in opened box.'' That coherence-breaking is the direct cause of the ``paradox,'' because the post-selection filter is orthogonal to the ``not seen'' state.

In their original analysis of the Three-box system, AV used the ABL retrodiction formula \citep{ABL}, which conditions the probabilities on the post-selection; this tends to hide from consideration the physically significant aspects of the post-selection. By describing the system in deceptively homely terms of boxes, the necessity of considering the effect of the measurement interaction is hidden --- looking into a box seems so simple and classical.  These approaches obscure the physical difference of the two observations (looking in one box or the other), allowing it to erroneously appear that the two boxes are holding the same particles with certainty in runs of the same experiment.  Thus ``paradox'' is suggested where there is none.

\section{Summary and conclusion}

In Appendix~\ref{A:ThreeBox} we describe what AV called ``a curious situation'' and has since been called the Three-Box ``paradox'': ``If a particle is certainly in box~1, then it is certainly not in box~2, and vice versa. Therefore, if a single particle is both certainly in box~1 and certainly in box~2 we have a paradox.''  But in the context of our careful description, the paradox is easily recognized as a mirage.%
\footnote{%
Also see \citet{Finkelstein06} for a nice clarification; his statement ``S2'' causes any sense of paradox in the Three-Box system to utterly disappear.
} %
Explicitly stating that the system interacts with two different detectors makes it obvious that the two certainties refer to distinct events, that the physical setting is different in the two cases, that symmetry makes the probabilities the same, and that it's merely a detail of the design (the specific choice of final state) that causes the common value of the probabilities to be 1.  

The Three-Box system is much more easily understood if expressed as a three-slit atomic diffraction system (\cf Appendix~\ref{A:ThreeSlit}). This form shows us that the Three-Box system's ``paradox'' is exactly the interesting but ordinary phenomenon of destructive interference in a two-slit Young apparatus, nothing more.

I made these points in my (\citeyear{Kirkpatrick03b}); they are sufficient to show that the Three-Box system is a matter of ordinary quantum mechanics, clever, but otherwise of no particular interest. To this, I added a classical system which exhibited the same ``paradoxical'' behavior regarding the certainty of each two disjoint events; this has the effect of showing that the so-called paradox is not even necessarily quantum-mechanical. 

\citet{RavonVaidman07} attack this classical example with several thrusts. They claim that the Three-Box system is a special ``quantum paradox,'' therefore not explicable nor implementable classically; however, they fail to establish a satisfactory meaning for this concept, and they fail to show either that the Three-Box system \emph{is} a quantum paradox or that my card game is \emph{not}. They suggest that my classical example violates scientific standards for explanatory models by having an \emph{ad hoc} dynamics implemented by measurement disturbance; however, they have set up a straw man simplification and do not actually deal with my system directly (and their argument is questionable even in the case of the simplified system).  And they do not deal with the fact that, by their own definition, my classical system successfully implements the ``paradox.''  

Thus I would repeat here what I said four years ago (\citeyear{Kirkpatrick03b}): 
\begin{quote}
The three-box example arose in a quantum setting, and was taken (somewhat uncritically) to be another example of the ``bizarre'' nature of quantum mechanics. The restatement of the example as a three-slit atomic Young system shows it to be a straightforward example of quantal interference. But exemplified in a perfectly ordinary setting, a deck of cards without a quantum in sight, the three-box phenomenon becomes merely an interesting phenomenon of ordinary probability systems, exhibited by a quantum mechanical system \emph{qua} probability system, in no way a quantal phenomenon --- hence [I would characterize comments regarding its bizarre nature] as ``much ado about nothing.''
\end{quote}

\appendix 
\def\appendixname{\hskip-1ex}
\setcounter{equation}{0}%

\bigskip\bigskip
\noindent\textbf{\large Appendices}

 \def\thesection{\Alph{section}}
 \def\thesubsection{\Alph{section}.\arabic{subsection}}
 \def\thesubsubsection{\Alph{section}.\arabic{subsection}.\arabic{subsubsection}}
 \def\theparagraph{\Alph{section}.\arabic{subsection}.\arabic{subsubsection}.%
                     \arabic{paragraph}}
 \def\thesubparagraph{\Alph{section}.\alph{subsection}.\arabic{subsubsection}.%
                     \arabic{paragraph}.\arabic{subparagraph}}
 \makeatletter
 \def\p@section{}
 \def\p@subsection{}
 \def\p@subsubsection{}
 \def\p@paragraph{}
 \def\p@subparagraph{}

 \def\@hangfrom@section#1#2#3{\@hangfrom{#1}{\textrm{#2}.~}{\textrm{#3}}}%
 \def\@hangfrom@subsection#1#2#3{\@hangfrom{#1}{\textrm{#2}}{\textrm{#3}}}%
 \def\@hangfrom@subsubsection#1#2#3{\@hangfrom{#1}{\textrm{#2}}{\textrm{#3}}}%
 \makeatother

\section{The Three-Box system}\label{A:ThreeBox}

The Three-Box system consists of a single particle and three boxes, labeled $j=1, 2,3$. The value $p_{j}$ of the observable $P$ denotes the presence of the particle in the corresponding box. The system is prepared in the state \ket{\psi} at the time $t_i$, and only those occurrences are considered for which, at time $t_f>t_i$, the system passes the filter \ket{\phi}; those initial and final states are 
\begin{equation}\label{E:threeboxQconditionEG}
 \ket{\psi}=(\ket{p_{1}}+\ket{p_{2}}+\ket{p_{3}})/\sqrt{3}\quad\text{and}\quad%
 \ket{\phi}=(\ket{p_{1}}+\ket{p_{2}}-\ket{p_{3}})/\sqrt{3}.
\end{equation}
If at the time $t$, $t_i<t<t_f$, we look in box 1, the state becomes a mixture of \ket{p_1} and $\ket{p_2}+\ket{p_3}$; the latter is orthogonal to \ket{\phi}. Therefore, for a particle that passes the filter \ket{\phi} at $t_f$, it is certain that, had we opened box~1 at $t$, the particle would have been found there.%
\footnote{\label{F:modify}%
That is, \emph{The state of the system is modified by the observation in such a way that whenever the box is empty, the post-selection becomes impossible}.
} %
But the system is symmetric in boxes~1 and 2, so: For every particle that ends up in state \ket{\phi} at $t_f$, had we instead opened box~2, the particle must certainly have been found  \emph{there}. The retrodictive observation of the particle being in box~1 is certain, and the retrodictive observation of the particle being in box~2 is certain.

\section{The Three Boxes as Three Slits (a quantum mechanical system)} \label{A:ThreeSlit}

Three slits are equally spaced with a separation $a$. The top and bottom slits are labeled 1 and 2, and the middle slit, 3. A detector $D$ is placed on-axis, at a distance $L$ from slit~3 so that $\sqrt{L^2+a^2}-L=\lambda/2$ (with $\lambda$ the wavelength); $a>>\lambda$. A detector $d$ is placed at slit~1 or slit~2 (for atoms, a micromaser, \emph{a la} \citet{ScullyWalther89}; for photons, a quarter-wave plate, the photon source linearly polarized). With this configuration, the initial and final states are as described by \eqref{E:threeboxQconditionEG} (with \ket{p_j} corresponding to passage through the $j$-th slit).

If $d$ is placed at slit~1, every detection at $D$ is in coincidence with a detection at $d$ (implying passage through slit~1); if $d$ is placed at slit~2, every detection at $D$ is in coincidence with a detection at $d$ (implying passage through slit~2). This behavior is easily understood: Placing a detector at slit~1 creates the disjunction ``either the particle passed through slit~1 or it passed through the double-slit apparatus comprised of slits~2 and~3''; but passage through the double-slit destructively interferes at the detector $D$, so the second term of the disjunction must be false, forcing the first to be true. This apparatus is symmetric under exchange of slits~1 and 2 --- hence the ``paradox.''

\bigskip
This three-slit system is exactly equivalent to the Three-Box example: The three slits correspond to the three boxes; a detector at slit~1 (only) corresponds to the opening of box~1 (only). The post-selection detector has been placed at the first minimum of the interference pattern of the two-slit (slits 2 and 3) Young apparatus, thus detecting only particles which were detected coming through slit 1.

 \renewcommand{\refname}{\textrm{\large References}}
 \footnotesize%

\begin{thebibliography}{}

\bibitem[Aharonov et~al.(1964)Aharonov, Bergmann, and Lebowitz]{ABL}
Aharonov, Y., Bergmann, P.~G., and Lebowitz, J.~L. (1964).
``Time symmetry in the quantum process of measurement,'' \emph{Phys. Rev.} {\bf
  134B}, 1410--1416.

\bibitem[Aharonov and Vaidman(1991)]{AharonovV91}
Aharonov, Y. and Vaidman, L. (1991).
``Complete description of a quantum system at a given time,'' \emph{J. Phys. A}
  {\bf 24}, 2315--2328.

\bibitem[Aharonov and Vaidman(2003)]{AharonovV03}
Aharonov, Y. and Vaidman, L. (2003).
``How one shutter can close $n$ slits,'' \emph{Phys. Rev. A} {\bf 67},
  042107--3, {arXiv:quant-ph/0206074}.

\bibitem[Davies(1976)]{Davies76}
Davies, E.~B. (1976).
{\em Quantum Theory of Open Systems}.
Academic Press, London.

\bibitem[Finkelstein(2006)]{Finkelstein06}
Finkelstein, J. (2006).
``What is paradoxical about the `{T}hree-box paradox'?,''
  {arXiv:quant-ph/0606218}.

\bibitem[Kastner(1999)]{Kastner99}
Kastner, R.~E. (1999).
``The {T}hree-{B}ox `paradox' and other reasons to reject the counterfactual
  usage of the {ABL} {R}ule,'' \emph{Found. Phys.} {\bf 29}, 851--863,
  {arXiv:quant-ph/9807037}.

\bibitem[Kastner(2004)]{Kastner04}
Kastner, R.~E. (2004).
``Shutters, boxes, but no paradoxes: {T}ime symmetry puzzles in quantum
  theory,'' \emph{International Studies in the Philosophy of Science} {\bf
  18}(1), 89--94, {arXiv:quant-ph/0207070}.

\bibitem[Kirkpatrick(2003a)]{Kirkpatrick03b}
Kirkpatrick, K.~A. (2003a).
``Classical {T}hree-{B}ox `paradox','' \emph{J. Phys. A} {\bf 36}(17),
  4891--4900, {arXiv:quant-ph/0207124}.

\bibitem[Kirkpatrick(2003b)]{Kirkpatrick03a}
Kirkpatrick, K.~A. (2003b).
``\,`{Q}uantal' behavior in classical probability,'' \emph{Found. Phys. Lett.}
  {\bf 16}(3), 199--224, {arXiv:quant-ph/0106072}.

\bibitem[Leifer and Spekkens(2005)]{LeiferSpekkens05}
Leifer, M.~S. and Spekkens, R.~W. (2005).
``Logical pre- and post-selection paradoxes, measurement disturbance and
  contextuality,'' \emph{Int. J. Theor. Phys.} {\bf 44}, 1977--1987,
  {arXiv:quant-ph/0412179}.

\bibitem[{L{\"u}ders}(1951)]{Luders51}
{L{\"u}ders}, G. (1951).
``{\"U}ber die {Z}ustands{\"a}nderung durch den {M}essprozess,'' \emph{Ann.
  Phys. (Leipzig)} {\bf 8}(6), 322--328.
English translation: Ann. Phys. (Leipzig) {\bf15} (9),
  663-670 (2006), arXiv:quant-ph/0403007.

\bibitem[Ravon and Vaidman(2007)]{RavonVaidman07}
Ravon, T. and Vaidman, L. (2007).
``The three-box paradox revisited,'' \emph{J. Phys. A} {\bf 40}(11),
  2873--2882, {arXiv:quant-ph/0606067}.

\bibitem[Scully and Walther(1989)]{ScullyWalther89}
Scully, M.~O. and Walther, H. (1989).
``Quantum optical test of observation and complementarity in quantum
  mechanics,'' \emph{Phys. Rev. A} {\bf 39}, 5229--5236.

\bibitem[Vaidman(1999a)]{Vaidman99a}
Vaidman, L. (1999a).
``Time-symmetrized counterfactuals in quantum theory,'' \emph{Found. Phys.}
  {\bf 29}, 755--765, {arXiv:quant-ph/9807075}.

\bibitem[Vaidman(1999b)]{Vaidman99b}
Vaidman, L. (1999b).
``The meaning of {E}lements of {R}eality---{R}eply to {K}astner,'' \emph{Found.
  Phys.} {\bf 29}, 865--876, {arXiv:quant-ph/9903095}.

\end{thebibliography}

\end{document}